\documentclass{aa}
\usepackage{graphicx}
\usepackage{amsmath}

\begin{document}

    \title{An Overdensity of Extremely Red Objects \\ Around Faint
    Mid-IR galaxies}   

   \author{P. V\"ais\"anen \inst{1,2}\thanks{Visiting Astronomer at
   the Infrared Telescope Facility, which is operated by the
   University of Hawaii under Cooperative Agreement no. NCC 5-538 with
   the National Aeronautics and Space Administration, Office of Space
   Science, Planetary Astronomy Program.} \and P.H. Johansson
   \inst{3,4}} 
   \offprints{P. V\"ais\"anen 
   \email{pvaisane@eso.org}}

\institute{European Southern Observatory, Casilla 19001, Santiago,
   Chile \and Departamento de Astronom\'ia, Universidad de
   Chile, Casilla 36-D, Santiago, Chile \and Observatory, P. O. Box
   14, FIN-00014 University of 
   Helsinki, Finland  \and Institute of Astronomy,
   Madingley Road, Cambridge, CB3 0HA, UK 
}

\date{Received / Accepted }
\authorrunning{V\"ais\"anen \& Johansson}
\titlerunning{ERO overdensity around mid-IR galaxies}

\abstract{We have searched for Extremely Red Objects (EROs) around
  faint mid-IR selected galaxies in ELAIS fields.  We find a
  significant 
  overdensity, by factors of 2 to 5, of these EROs compared to field
  EROs in the same region and literature random field ERO
  counts. The excess is similar to that found previously in the fields
  of known high redshift quasars and AGN.  While with the present
  data it cannot be definitely shown whether the overdensity is
  physically connected to the mid-IR source, a derived radial
  distribution does suggest this.  The 
  fraction of EROs among $K$-selected galaxies is high in the mid-IR
  fields in agreement with the picture that the EROs responsible for
  the overdensity are members of high redshift clusters of
  galaxies.  We find $R-K>5$ selected EROs to be 
  more clustered around the mid-IR galaxies than $I-K>4$ EROs,
  though statistics are weak because of small numbers.  However, this
  would be consistent with a cluster/galaxy group scenario if, as
  we argue, the former colour selection finds preferentially more
  strongly clustered early type 
  galaxies, whereas the latter selection includes a larger fraction of
  dusty EROs.  Finally, using the mid-IR data, we are able to limit
  the fraction of ULIRG type very dusty EROs at $K<18$ magnitude to
  less than $\sim10$\% of the total ERO population.

}

\maketitle

\section{Introduction}

Galaxy formation and evolution is undoubtedly one of the major issues
under debate in modern astronomy and cosmology.  On the observational
side, there has been a great effort to gather samples of galaxies
against which the successes and short-comings of galaxy 
formation models can be weighed.  Two useful approaches are
determining the space densities and ages of the most massive galaxies
at and beyond $z\sim1$, and identifying clusters or
proto-clusters of galaxies at those redshifts. 

The definition of Extremely Red Objects (EROs) for example by colours
of $R-K>5, I-K>4$, is motivated by this selection pin-pointing the
crucial evolved and most massive galaxies at $z>1$ from optical and
near-infrared extragalactic imaging surveys.  
The models are often presented as two extremes, where on one
side there are monolithic collapse models (e.g.\ 
Eggen, Lynden-Bell \& Sandage \cite{Eg62}; Larson \cite{La75})
with pure luminosity evolution, 
and on the other hierarchical assembly scenarios (e.g.\ White \& Rees
\cite{Wh78}; White \& Frenk \cite{Wh91}; Somerville \& Primack
\cite{So99}; Cole et al. \cite{Co00}).  The predicted numbers of
evolved galaxies at high 
redshift have been very different, at least 
until recently (see eg.\ Somerville et al.\ \cite{So04}), and thus in
principle easy to test against observations.  However, in addition to
the fact that reality possibly is some mix of models (e.g.\ Roche
et al.\ \cite{Ro02}), the results from observed numbers of high-$z$
ellipticals have also been inconclusive.

There are two difficulties in interpreting the ERO number count
results.  First of all, the extreme red colours can be
attributed also to dust reddened starburst galaxies 
thus contaminating the old distant early type samples (see e.g.\
recent papers by Cimatti et al. \cite{Ci03}; Yan \& Thompson 
\cite{Ya03}; Takata et al. \cite{Ta03}; and
references therein).  According to
Cimatti et al. (\cite{Ci02}) and Smail et al.\ (\cite{Sm02})
$30-60$\% of EROs at typical ERO brightnesses of $K\sim18-20$ might be
dusty.  At brighter levels, $K<17.5$, we find a smaller percentage
$<20$\% of dusty EROs (V\"ais\"anen \& Johansson 2004, hereafter
Paper~I).  Recent HST morphological studies have
complicated the picture somewhat by finding large fractions of disk
galaxy EROs (Yan \& Thompson \cite{Ya03}; Moustakas et
al.\ \cite{Mo04}; Gilbank et al.\ \cite{Gi03}).  In addition to
morphological, spectroscopic, or colour property methods, the
separation of dusty and early type EROs
can also be done using detections at longer wavelengths, where
ellipticals are not expected to radiate much.  For attempts in sub-mm
to radio see e.g.\ Mohan et al. (\cite{Mo02}), Smail et al.\
(\cite{Sm02}), Wehner, Barger, \& Kneib (\cite{We02}), Webb et al.\
(\cite{We04}). In Paper~I we used for the first time a mid-IR data-set
to select dusty EROs -- the difference of ellipticals and starforming
galaxies is very large also in the mid-IR, and e.g.\ the
identification problems of large sub-mm beams can be avoided.
However, the exact relative fractions of different types of EROs as a
function of magnitude is still an open question. 

The second reason for varying numbers of EROs is their strong
clustering (Daddi et al.\ \cite{Da00}; Firth et al.\ \cite{Fi02};
Roche et al.\ \cite{Ro02}), which is in fact expected if EROs are
the counterparts of present day massive ellipticals.  This
clustering ties the ERO studies together with the second approach
to understanding galaxy and large scale structure formation,
namely the search for high redshift clusters of galaxies (see e.g.\ 
Olsen et al.\ \cite{Ol99}; Gonzalez et al. \cite{Go01}; Bahcall et
al.\ \cite{Ba03}; Mullis et al.\ \cite{Mu03}).  
It is feasible that EROs are found in clusters
and concentrations of EROs can be used to locate high-$z$ clusters of
galaxies. 

Significant overdensities of EROs, compared to large field
surveys, have been found in high-$z$ AGN and QSO fields (e.g.\
McCarthy, Persson, \& West \cite{Mc92}; Hu \& Ridgeway \cite{Hu94};
Dey, Spinrad, \& Dickinson \cite{De95}, Cimatti et al.\ \cite{Ci00};
Chapman, McCarthy \& Persson \cite{Cha00};  Hall et al.\ \cite{Ha01};
S\'anchez \& Gonz\'alez-Serrano \cite{Sa02}; Best et al.\ \cite{Be03};
Wold et al.\ \cite{Wo03}).
As discussed in these studies, it is not yet clear whether the
overdensity is necessarily physically connected to the AGN:  
it is also possible that the EROs would be part of foreground 
overdensities of galaxies which enhance the probability of detection
of these AGN via gravitational lensing (see e.g.\ discussion in Wold et
al. \cite{Wo03}).  Nevertheless, in either case the
EROs can be used to identify candidates for high redshift galaxy
clusters, and thus once again present themselves as an important window
into the galaxy and structure formation era of the Universe.

We present here a NIR survey performed in the European Large Area
  ISO 
Survey (ELAIS; Oliver et al.\ \cite{Ol00}) fields, and derive surface
densities of EROs.  Rather than a
survey with large continuous sky coverage (as in Paper~I) the 
deeper NIR data-set of the present paper consists of numerous
  individual  
pointings around faint ELAIS ISOCAM detections or blank fields.  We
  make 
use of the newly available Final band merged ELAIS Catalogue
(Rowan-Robinson et al. \cite{Row04}; Vaccari et al.\ 2004)
and a publicly available optical survey.  

In particular, as the main aim of this paper, we compare the
distribution of EROs in ``random'' fields to those fields which
are occupied by faint mid-IR detections.  We are thus able to determine
whether there are similar overdensity biases of EROs as in the case of
verified high-$z$ AGN.  Similar results are expected at least to some
extent, since a significant fraction of faint ISOCAM sources appear to
be  
AGN at $z>1$ (e.g.\ Fadda et al. \cite{Fa02}; La Franca et
al. \cite{La04}; Johansson, V\"ais\"anen \& Vaccari 2004, in prep.).   
As mentioned, locations of ERO overdensities may serve as candidates
for  
high-redshift clusters and groups of galaxies, whose history in turn
place 
strong constraints on structure formation scenarios.  A demonstrated
link 
between ERO concentrations and specific classes of sources (in this
case AGN or certain mid-IR galaxies) would then further facilitate
finding the crucial clusters using data such as large MIR surveys
with the Spitzer Space Telescope. 

The structure of this paper is as follows. In Sect.\ 2 we present the
observational data.  In Sect.\ 3 the properties
of the ERO samples are defined and 
analysed, and Sect.\ 4 discusses the overdensity of EROs.
We assume throughout this  
paper a flat ($\Omega_{0}=1$) cosmology with $\Omega_{m}=0.3$,
$\Omega_{\Lambda}=0.7$ and $H_{0}=70 \ \rm km s^{-1} Mpc^{-1}$.

\section{Observations}

\subsection{ISO data}

The \textit{ISO} data used in this study comes from the Final
band-merged ELAIS catalogue (Rowan-Robinson et al.\
\cite{Row04}; Vaccari et al.\ 2004).  For further
description of the ELAIS project, 
observations, data reduction, source extraction and extragalactic
source counts see Oliver   
et al. (\cite{Ol00}), Serjeant et al. (\cite{Se00}), and Efstathiou et
al. (\cite{Ef00}). 
The observations used in this study were primarily made using 
the ISOCAM (Cesarsky et al. \cite{Ce96}) LW3 filter at 15 $\mu$m.

\subsection{Near-Infrared data}

The observations presented here were carried out with the 
3.0-m NASA Infrared Telescope Facility (IRTF) on Mauna Kea, and
were conducted in the ELAIS fields N1 and 
N2, centered at $(\alpha,\delta)=16\rm h\, 09\rm m\, 00\rm s,
54^{\circ}\, 40'\, 00\arcsec$ and $(\alpha,\delta)=16\rm h\, 36\rm m
\, 00\rm s,
41^{\circ}\, 06'\, 00\arcsec$ J2000.0 respectively.   
The data were acquired during five photometric
nights in June 1998, and the seeing varied between 
$0.7\arcsec-1.0\arcsec$.  
The data set consists of images taken at
(very preliminary) ISOCAM detection positions where no obvious 
counterparts were found in the Digital Sky Survey (DSS).
The sample used here consists of a total of 33 images. 
Of these, 9 were taken from the ELAIS N1 region and 24 from the ELAIS
N2 region.  The total surveyed area is $73.6 \ \rm arcmin^{2}$.
These areas are small non-continuous regions of the entire ELAIS N1
and N2 fields,  which cover a continuous area of $> 2 \ \rm deg^{2}$. 

We used the NFSCam instrument containing a $256 \times 256$ format
InSb detector array.  A $0.3\arcsec$ pixel scale was used for all
observations.
Images were constructed from co-added dithered frames 
using standard NIR data reduction methods (such as in DIMSUM) written
as IDL procedures.  The final 
images have a field of view of $1.5' \times 1.5'$.
The total integration 
times were in the range of 20 -- 30 minutes and the limiting magnitude
for the deepest images was around $K=20.0$.

\subsection{Optical data} 
\label{optdata}

Optical photometric data were obtained from the Isaac Newton
Telescope (INT) Wide Field Survey (WFS).  
The WFS data are publicly available on the Cambridge Astronomical
Survey Unit (CASU)  
homepage\footnote{http://archive.ast.cam.ac.uk/}. See McMahon et
al. (\cite{Mc01}) for a review of the INT Wide Field Survey Project. 
  We use data consisting of Vega-based $g',r',i',Z$ band photometry 
  (the apostrophes are dropped henceforth). The filters are similar 
to the SDSS (Sloan Digital Sky Survey) filters (Fukugita et
al. \cite{Fu96}).  We heavily rely on the $r,i$ bands, and use the
following filter transformations to the standard Cousins system
(Landolt \cite{La92}; WFS
webpage\footnote{http://www.ast.cam.ac.uk/$\sim$wfcsur/index.php}):  
$r-R = 0.275 (R-I) + 0.008$ and $i-I=0.211(R-I)$.
Our adopted $5\sigma$ detection limits for a $1\arcsec$ seeing are
$g=25.0$, $r= 24.1$, $i=23.2$~ and $z=22.0$ (WFS webpage).
For further details on the 
pipeline processing of INT wide field survey data consult 
Irwin \& Lewis (\cite{Ir01}) and Gonzalez-Solares et
al. (\cite{Go04}) for discussion of the WFS data in ELAIS fields in 
particular.

\subsection{Other data}

In addition to 90 and 175 $\mu$m, and 20 cm data available in the
ELAIS catalogue, we checked several other separate surveys in the same 
region.  A region of the ELAIS N2 field has been surveyed in the
sub-millimetre with SCUBA to a $3.5\sigma$ limit $F(850\mu m)\simeq 8
\rm \ mJy$ by Scott et al. 
(\cite{Sc02}). In the x-ray, the \textit{ROSAT} (0.1-2 keV) all-sky
survey was cross-correlated with ELAIS sources (Basilakos et
al. \cite{Ba02}) and deep 75 ks Chandra observations were performed in
N1 and N2 regions (Almaini et al. \cite{Al03}; Manners et
al.\ \cite{Ma03}).  None of these surveys produced overlapping sources
with the final list of EROs analyzed in this paper.  In the same N2
area Roche et al. (\cite{Ro02}) have performed a deeper study of the
clustering, number counts and morphology of EROs. Only one of our IRTF
images overlaps with this very well studied region.

\subsection{Astrometry and photometry}

The final astrometry of the IRTF data was obtained by
matching the visible sources with 
with Hubble Guide Star Catalog 2 objects (GSC2, available at the GSC 
homepage\footnote{http://www-gsss.stsci.edu/gsc/GSChome.htm})
giving us an accuracy of $\approx0.3\arcsec$.
The INT astrometry was derived using GSC objects, which results in an
external astrometric accuracy of $0.5\arcsec-1.0\arcsec$ (Irwin \&
Lewis \cite{Ir01}). 

The photometry and calibration of the optical INT Wide Field Survey
data were obtained from the archives (Irwin \& Lewis \cite{Ir01}; 
Gonzalez-Solares et al. \cite{Go04}).

All our near-IR photometry is performed using the SExtractor software
(v.2.2.1 and v.2.3; Bertin \& Arnouts \cite{Be96}). 
For detecting objects in the IRTF data we used the following
SExtractor settings: a 1.25$\sigma_{\rm sky}$ 
(translating to $K\sim20.7$ mag $\rm arcsec^{-2}$ 
for integration times $\ge 20$ min and $K\sim20.4$ mag $\rm
arcsec^{-2}$ for those with 10--15 min) detection threshold in a
minimum area of 5 
pixels ($0.45 \rm \ arcsec^{2}$), with a detection filter of 2.5
pixels Gaussian FWHM, approximately matching the seeing.  

For total magnitudes, the SExtractor BEST-magnitude was used -- it is
most often the Kron-magnitude (Kron \cite{Kr80}), where an elliptical
aperture for 
photometry is defined by the shape and size of the detected object.
We found these magnitudes to be robust over 
a wide range of magnitudes and source profiles and
contain (nearly) all of the object flux.
In very crowded regions the BEST-magnitude can also be the 
isophotal magnitude instead.  
However, since the WFS data is given in 2.4\arcsec \ diameter
apertures, we 
calculated corresponding aperture magnitudes from the IRTF 
images.  All optical-near-IR colours given in this paper are thus
calculated using matched apertures.

As mentioned, the deepest parts of the NIR data reach $K=20$.
Noticeable incompleteness starts affecting NIR source counts at $K>19$
mag.  Since we are interested in the very reddest sources, and the
detection limits of the optical bands are $r\approx24.1, \,
i\approx23.2$ 
we will not consider NIR sources fainter than $K=18.7$ in this
paper. We thus do not need any completeness corrections to number
count data.

The magnitude calibration was determined from short exposures of UKIRT
faint standard stars 
and HST/NICMOS IR standards. From these observations a magnitude
zeropoint of 21.47 was derived,
which is consistent with the IRTF/NFSCAM manual value. We estimate our
$K$-band photometry to be accurate within 0.06 mag.  We also note that
the ERO list in Roche et al.\ ({\cite{Ro02}) has 4 objects within one
of our IRTF frames. Of these, 3 have fully consistent magnitudes
though the fourth is 0.7 mag brighter in our catalogue.

\section{Construction of the ERO samples}
\label{ero-samples}

\subsection{Definition of EROs}
\label{definition}

Numerous different selection criteria have been defined for EROs,
including $R-K\ge 6$, $R-K\ge5.3$, $R-K\ge5$, $I-K\ge4$ 
with $K$-magnitude upper limits from 18 to 20. 
In this paper we use the following definition for
EROs: $r-K\ge 5.5$ and/or $i-K\ge 4.4$. Using the
colour transformations given in Sec.~\ref{optdata} above, and typical
colours of our objects, $R-I\approx 1.4-2$ (see e.g.\ Fig.~2 of Paper
I), it is seen that our choices correspond to the widely used
$R-K\ge5$, $I-K\ge4$ ERO selection criteria.
All ERO selections are designed for selecting early type galaxies at
$z\ge 1$ -- however, we note that 
since the model colours that different authors use vary, 
there might easily be 0.1-0.3 mag differences in the colours at
$z\sim1$.  The elliptical galaxy models we used (GRASIL library; Silva
et al. \cite{Si98}) become EROs at $z\approx1.1-1.2$ with 
the criteria defined above.

Figure~\ref{silva_model} shows $r-K$ and $i-K$ model colours of
several  
representative galaxies against redshift, along with the ERO criteria.
Model SEDs are adopted from the GRASIL library 
(Silva et al. \cite{Si98}; a sample set of model SEDs and procedures
to calculate more are available
online\footnote{http://web.pd.astro.it/granato/}). 
Ordinary spirals never reach the red colours of EROs, while both 
ellipticals and reddened starbursts become EROs
when seen beyond $z\sim1$.  For comparison, the colour of the
prototype dusty ERO HR10 is  also plotted as a function of
redshift. 
The lowest panel shows the flux ratio $f_{15 \mu m}/f_{2.2 \mu m}$ -- 
the degeneracy in the red colours of old 
ellipticals and dusty starbursts is clearly broken.

   \begin{figure}
\resizebox{8.0cm}{!}{\includegraphics{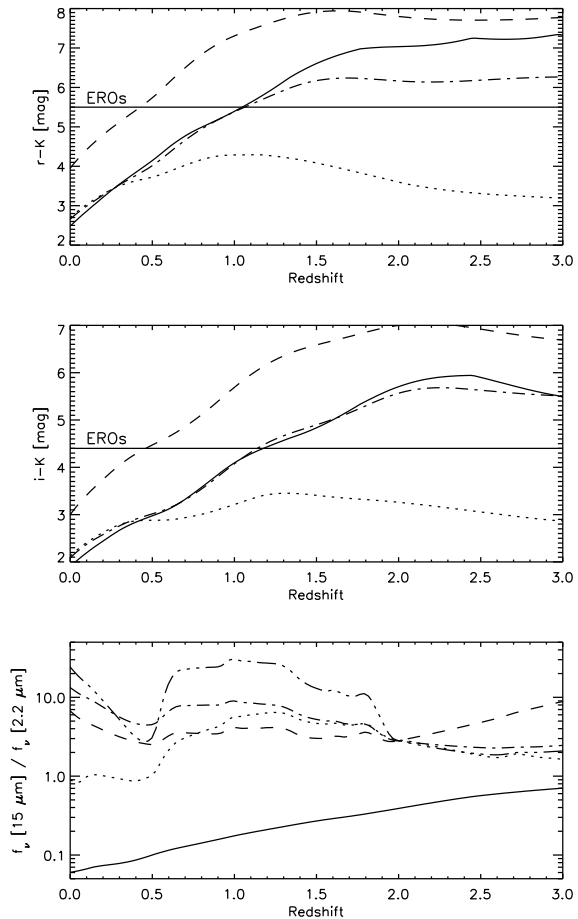}}
      \caption[]{Observed $r-K$ and $i-K$ and MIR/NIR colours as a
	function of redshift. Representative model galaxy SEDs are
	taken  
from GRASIL models (Silva et al. \cite{Si98}): an 
evolving Sb spiral (dotted line), an evolving elliptical (solid
	curve), non-evolving prototypical starburst M82 (dash-dot),
	and HR10 ERO (dashed).  The dash-triple-dot curve
	in the lowest panel corresponds to the ULIRG ARP220. The
	horizontal lines show the colour definition of EROs in this
	paper.}  
         \label{silva_model}
   \end{figure}

\subsection{Matching of optical and NIR data}

The optical counterparts of near-IR sources were extracted from the
INT WFS catalogue using a 1.5\arcsec \ search radius. Since the WFS
catalogue contains multiple detections of the same source (due to 
overlapping CCD frames), optical counterparts within 0.5\arcsec \ were
averaged.  After the purging, the brightest optical source of any 
remaining multiple counterparts was selected.
There were $118/722=16\%$ NIR sources which were undetected in the WFS
catalogue. This is expected, since the IRTF survey goes somewhat deeper
than the WFS data.  For undetected sources
the respective $5\sigma$ limiting magnitude was assigned as the
limiting brightness. 

\subsection{Star vs.\ galaxy separation}
\label{separation}

Next, stars were separated from galaxies.  Stellarity indices are
available from both the NIR catalogue (the SExtractor CLASS parameter)
and the WFS catalogue, where a flag defines galaxies, definite stars,
and 
various degrees of uncertain stellarities.  Since the IRTF data is
deeper the NIR data provide more reliable classifications than the
WFS. 

For the EROs we are interested in, the reddest and faintest objects in
the lists, we found that a colour selection works at least as well, if
not better, than the morphological separation.  Please refer to Sect.\ 
3.3 and Fig.\ 2 in Paper~I, where a separating line of $r-K >
2.16(r-i)+1.35$ for galaxies is derived.  At the $r-K>5$ regime there
is no overlap 
between the populations.  In essence, very late type cool stars (in
particular of spectral type L) can have optical to NIR colours very
similar to extragalactic EROs (e.g.\ Dahn et al. \cite{Dah02}).
However, the $r-i$ colour is $>2$ for 
these stars, whereas galaxies are bluer.  Using this colour, extremely
red stars can be discarded from the ERO samples.  

In practice, we used a combination of methods:  At $K<18$
we define the stellar sources according to NIR morphology (those with
CLASS$>$0.5).  At $K>18$ those with CLASS$>$0.5 {\em and} a stellar
colour 
are assigned as stars (if detected optically, only morphology used if
no optical detection was available). 
We find 128 stars at $K<18$ mag, which is in very good agreement with
the prediction of 115 stars for these fields from the Galactic
foreground point source model SKY (Cohen et al.\ \cite{Co94}; see also
V\"ais\"anen et al.\ \cite{Va00}). 
As for the colours of stars, approximately 55\% and 35\% of objects
redder than $r-K>5.5$ (i.e.\ $R-K>5$ are found to be stars at the
limits $K<18$ and $K<18.5$, 
respectively.  These are very consistent with respective values
calculated from the stellar distribution model of Jarrett 
(\cite{Ja92}; Wold, private communication; Wold et al.\ \cite{Wo03}). 
In summary, it should be noted that in contrast to fainter $K>19$ ERO
surveys, brighter ERO surveys are significantly contaminated by very
cool stars in our own Galaxy.

\subsection{Extraction of EROs}

We searched for EROs from the resulting galaxy catalogue  
according to the colour definitions given above in
Section~\ref{definition}. Only NIR detections at $5\sigma$ level and
over were considered.  As already mentioned, there were 118 sources 
which were not detected to the WFS survey limits: These are assigned
magnitudes of $r>24.1$ and $i>23.2$, and thus may enter our ERO
catalogue with colours $r-K>5.5$ and $i-K>4.4$.  We then went  
through the resulting list checking our NIR maps as well as the WFS CCD
images for any obvious spurious objects. 
Ultimately there were 27 EROs in the matched IRTF catalogue using the
$r-K$ criterion and 27 using the $i-K$ limit --- 14 EROs are common
to lists resulting from both selection criteria, i.e.\ there were 40
EROs found in total (Table~\ref{Tab0-phot}).  Note in particular
  that 19/40 of the EROs have only upper limits in optical bands, ie.\
  come from among those 118 sources detected only in the NIR.  An
  example of an ERO field is shown in Fig.~\ref{eroexample}. 

   \begin{figure}
\resizebox{8.5cm}{!}{\includegraphics{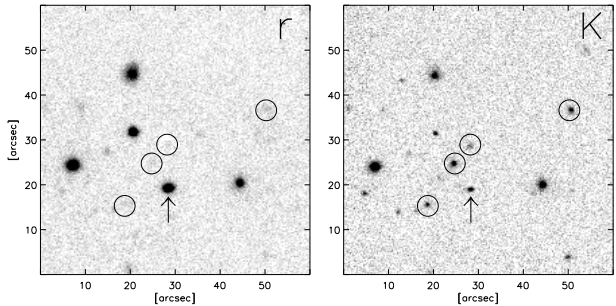}}
      \caption[]{An example of one of our IRTF $K$-band fields; the
	$r$-band data is from the WFS archive.  There are many red
	sources in this field around ELAISC15\_J163541.7+405900
	(indicated by the arrow), and those satisfying our ERO
	criteria are circled.
	The four sources correspond to EROs \#7, 6, 34, and 23 in
	Table~\ref{Tab0-phot}, starting from the lower left.  The
	extremely red ERO-6 lies within $3\sigma$ of positional
	uncertainty of the MIR source (see Section~\ref{dustysep}).
}
         \label{eroexample}
   \end{figure}

The total numbers of 
verified EROs are summarized in Table \ref{tab1}, along
with some other survey characteristics.  Colour-magnitude plots
are shown in Fig.~\ref{irtf_comb}.  We also note that of the 4 common
EROs with the Roche et al.\ (\cite{Ro02}) catalogue, one is excluded
from our survey as a star and another because of detection levels
(i.e.\ it is not detected in the WFS, but a $K>18.7$ magnitude does
not necessarily make it an ERO as far as our data is concerned). 

   \begin{figure*}
\resizebox{18cm}{!}{\includegraphics{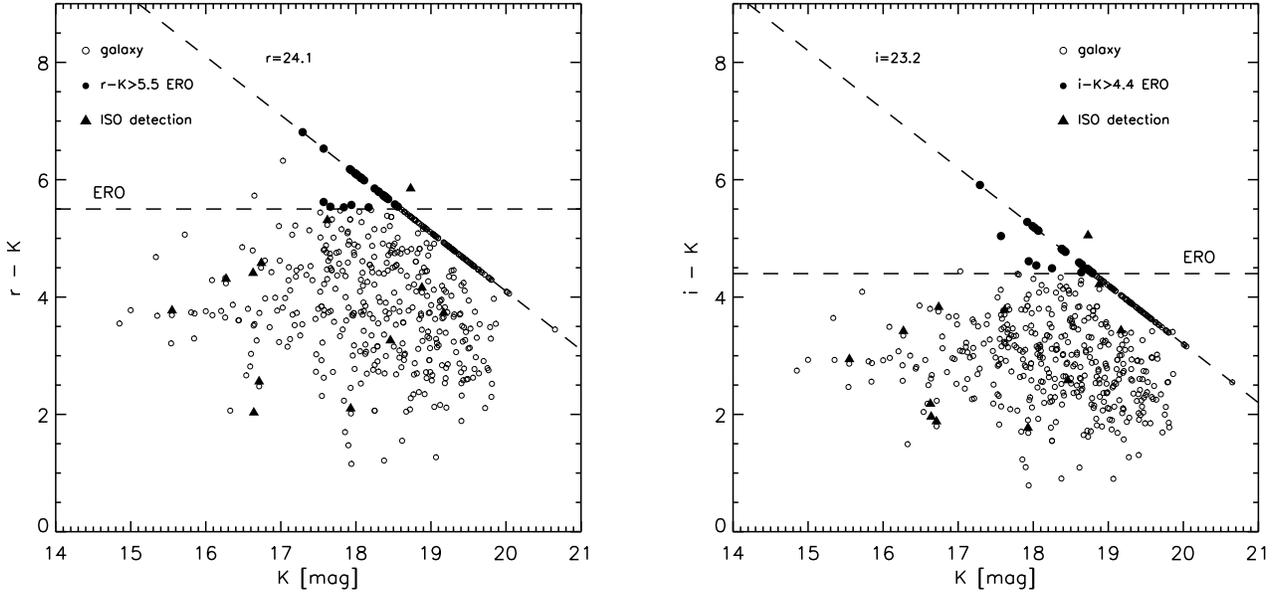}}
      \caption[]{The $K$-band magnitude vs. the $r-K$~and $i-K$
	colours.  Only galaxies are plotted.  Verified EROs are
overlaid with solid circles, and ISOCAM detections as solid triangles
The objects lying on the $r=24.1$~and $i=23.2$ lines are sources which
	were not 
detected in the WFS and for which the 5$\sigma$  
detection limit is assigned.  Note that the total $K$-magnitude is
used on the x-axis, whereas the colour is calculated with matching
small apertures.}
         \label{irtf_comb}
   \end{figure*}

   \begin{table*}
\begin{center}
\begin{tabular}{lccccccccccccc}
     \hline
     \hline
     \noalign{\smallskip}
  & $N$  & $N_{\star}$ & Area $\rm arcmin^{2}$ & $K_{\rm lim}$ & $<r-K>$ & $<i-K>$ & \multicolumn{3}{c}{EROs} & \multicolumn{2}{c}{$\rm ERO/arcmin^{2}$} \\
(1)      & (2) & (3) & (4) & (5) & (6) & (7) & (8) & (9) & (10) & (11)   & (12) \\
            \noalign{\smallskip}
            \hline
            \noalign{\smallskip}
IRTF       & 722 & 128 & 74 & 19.5 & 3.80 & 2.90 & 27 & 27 & 14 & 0.36 & 0.122--0.189 \\
\noalign{\smallskip}
\hline
\end{tabular}
\end{center}
\caption[]{Columns (2) to (5) give the total
  numbers of objects and stellar sources in our survey, the covered
  area, and the characteristic
  limiting magnitude. The rest of the columns refer to
  galaxies only: columns (6) to (7) give mean colours, though note
  that these are lower limits since optical non-detections we ignored.
  Column (8) gives the  number of EROs selected by
  $r-K>5.5$, (9) by $i-K>4.4$, and (10) by their combination; column
  (11) is the surface density corresponding to columns (8) and (9),
  and (12) is the range of 
  surface densities at $K<18$ resulting from either $r$ or  
$i$ band based selection.}  
\label{tab1}
\end{table*}

  \begin{table*}[b]
\footnotesize
\begin{center}
\begin{tabular}{cccrrrrr}
     \hline
     \hline
     \noalign{\smallskip}
  &  RA (J2000) & DEC (J2000) & $r'$ & $i'$ & $Z$ & $K$  \\
            \noalign{\smallskip}
            \hline
   1 & 242.275418 &  54.967102 &  $>24.10  $ & $>23.20 $ &  $>22.00 $ &  $18.38 \pm   0.11$  &  \\
   2 & 242.365120 &  54.469628 &  $>24.10  $ & $>23.20 $ &  $>22.00 $ &  $18.43 \pm   0.10$&\\
   3 & 248.588973 &  41.052093 &  $>24.10  $ & $>23.20 $ &  $21.09 \pm 0.09$ &  $18.06 \pm   0.08$&\\
   4 & 248.593709 &  41.053258 &  $>24.10  $ & $>23.20 $ &  $>22.00 $ &  $18.07 \pm   0.08$&\\
   5 & 248.597116 &  41.044969 &  $>24.10  $ & $>23.20 $ &  $>22.00 $ &  $17.99 \pm   0.09$&\\
   6 & 248.925210 &  40.985268 &  $>24.10  $ & $22.61 \pm 0.15$ &  $21.38 \pm 0.12$ &  $17.57 \pm   0.07$&\\
   7 & 248.927333 &  40.982732 &  $>24.10  $ & $22.74 \pm 0.16$ &  $>22.00 $ &  $18.25 \pm   0.13$&\\
   8 & 248.931334 &  41.136512 &  $23.51 \pm  0.12 $ & $22.55 \pm 0.14$ &  $>22.00 $ &  $17.94 \pm   0.08$&\\
   9 & 248.964062 &  41.101228 &  $>24.10  $ & $>23.20 $ &  $>22.00 $ &  $18.02 \pm   0.10$&\\
  10 & 248.994868 &  41.063800 &  $>24.10  $ & $>23.20 $ &  $21.57 \pm 0.14$ &  $18.38 \pm   0.11$&\\
  11 & 249.022827 &  40.783118 &  $>24.10  $ & $>23.20 $ &  $>22.00 $ &  $17.29 \pm   0.06$&\\
  12 & 249.059398 &  40.794446 &  $>24.10  $ & $>23.20 $ &  $>22.00 $ &  $17.92 \pm   0.17$&\\
  13 & 249.222717 &  40.992068 &  $>24.10  $ & $>23.20 $ &  $>22.00 $ &  $18.41 \pm   0.24$&\\
  14 & 249.311503 &  41.158857 &  $>24.10  $ & $>23.20 $ &  $21.52 \pm 0.14$ &  $18.39 \pm   0.12$&\\
            \noalign{\smallskip}
\hline
            \noalign{\smallskip}
  15 & 241.931449 &  55.037148 &  $>24.10  $ & $22.54 \pm 0.10$ &  $>22.00$ &  $18.40 \pm   0.13$&\\
  16 & 242.382161 &  54.471598 &  $>24.10  $ & $22.38 \pm 0.11$ &  $21.27 \pm 0.09$ &  $18.30 \pm   0.10$&\\
  17 & 242.984000 &  54.705978 &  $>24.10  $ & $22.62 \pm 0.13$ &  $21.13 \pm 0.09$ &  $18.56 \pm   0.15$&\\
  18 & 248.598040 &  40.953026 &  $>24.10  $ & $22.16 \pm 0.10$ &  $20.92 \pm 0.08$ &  $18.08 \pm   0.11$&\\
  19 & 248.655921 &  40.951751 &  $23.20 \pm  0.09 $ & $21.50 \pm 0.05$ &  $20.53 \pm 0.05$ &  $17.66 \pm   0.09$&\\
  20 & 248.870571 &  41.003286 &  $>24.10  $ & $22.07 \pm 0.09$ &  $21.30 \pm 0.10$ &  $18.52 \pm   0.15$&\\
  21 & 248.880417 &  41.007793 &  $>24.10  $ & $21.98 \pm 0.09$ &  $20.69 \pm 0.07$ &  $17.94 \pm   0.09$&\\
  22 & 248.880647 &  41.005380 &  $>24.10  $ & $22.32 \pm 0.11$ &  $21.15 \pm 0.09$ &  $18.11 \pm   0.10$&\\
  23 & 248.915574 &  40.988594 &  $23.37 \pm  0.11 $ & $21.63 \pm 0.06$ &  $20.89 \pm 0.07$ &  $17.84 \pm   0.08$&\\
  24 & 248.936320 &  41.151117 &  $23.70 \pm  0.14 $ & $22.29 \pm 0.11$ &  $>22.00 $ &  $18.17 \pm   0.11$&\\
  25 & 249.062402 &  40.825955 &  $23.19 \pm  0.10 $ & $21.69 \pm 0.07$ &  $20.77 \pm 0.07$ &  $17.57 \pm   0.06$&\\
  26 & 249.311868 &  41.159109 &  $>24.10  $ & $22.70 \pm 0.16$ &  $>22.00 $ &  $18.36 \pm   0.12$&\\
  27 & 249.349512 &  41.090600 &  $>24.10  $ & $22.31 \pm 0.12$ &  $>22.00 $ &  $18.00 \pm   0.07$&\\
            \noalign{\smallskip}
\hline
            \noalign{\smallskip}
  28 & 242.362595 &  55.183170 &  $23.53 \pm  0.14 $ & $22.58 \pm 0.09$ &  $20.68 \pm 0.07$ &  $18.04 \pm   0.07$&\\
  29 & 242.382952 &  54.472045 &  $>24.10  $ & $>23.20 $ &  $>22.00 $ &  $18.79 \pm   0.13$&\\
  30 & 242.570354 &  55.116371 &  $>24.10  $ & $23.06 \pm 0.14$ &  $21.56 \pm 0.14$ &  $18.64 \pm   0.16$&\\
  31 & 248.587252 &  40.971250 &  $>24.10  $ & $>23.20 $ &  $>22.00 $ &  $18.75 \pm   0.16$&\\
  32 & 248.589808 &  41.050045 &  $>24.10  $ & $>23.20 $ &  $>22.00 $ &  $18.72 \pm   0.12$&\\
  33 & 248.884767 &  41.009763 &  $>24.10  $ & $>23.20 $ &  $>22.00 $ &  $18.64 \pm   0.14$&\\
  34 & 248.923854 &  40.986371 &  $>24.10  $ & $>23.20 $ &  $>22.00 $ &  $18.65 \pm   0.14$&\\
  35 & 248.929511 &  41.137141 &  $>24.10  $ & $>23.20 $ &  $>22.00 $ &  $18.72 \pm   0.13$&\\
  36 & 248.930122 &  41.131120 &  $>24.10  $ & $>23.20 $ &  $>22.00 $ &  $18.73 \pm   0.16$&\\
  37 & 249.024081 &  40.781571 &  $>24.10  $ & $>23.20 $ &  $>22.00 $ &  $18.61 \pm   0.13$&\\
  38 & 249.047083 &  40.814442 &  $>24.10  $ & $>23.20 $ &  $>22.00 $ &  $18.78 \pm   0.14$&\\
  39 & 249.103560 &  41.404861 &  $>24.10  $ & $>23.20 $ &  $>22.00 $ &  $18.79 \pm   0.16$&\\
  40 & 249.234015 &  40.996663 &  $>24.10  $ & $>23.20 $ &  $>22.00 $ &  $18.75 \pm   0.32$&\\
\noalign{\smallskip}
\hline
\end{tabular}
\end{center}
\caption[]{Photometry of the found EROs.  Sources 1--14 are those
  with $r-K>5.5$ and $i-K>4.4$, while 15--27 and 28--40 are selected
  by only $r-K>5.5$ and $i-K<4.4$, respectively.}
\label{Tab0-phot}
\end{table*}

\section{Discussion}

\subsection{Separation of dusty and early type EROs}
\label{dustysep}

One of the motivations of this study is to attempt the se-pa\-ration
of dusty EROs from the evolved early type population using
mid-IR data.  
As seen in the lowest panel of Fig.~\ref{silva_model} the $f_{15 \mu
  m}/f_{2.2 \mu m}$ flux ratio is below unity for  
an elliptical galaxy in the redshift range of $z=0-3$, whereas the
ratio is $\sim 10$ for a typical dusty ERO.  The $f_{15 \mu m}/f_{2.2
  \mu m}$ ratio is thus in principle a very powerful discriminator
between dustfree and dusty EROs. 

What fraction of the EROs, to a given flux limit, are detected in the
mid-IR?  Because our survey is not a pure field survey, but rather a
targeted one, we cannot answer the question in general. Various
limits can be obtained, however.  

One clear ISO-ERO, and two ambiguous cases were detected among
the total of 40 IRTF EROs.  In one of the ambiguous cases, an ISOCAM
source lies within 3\arcsec \ of both an ERO and a spectroscopically
confirmed non-ERO Sy1 galaxy at $z=1.150$ which is selected as the
counterpart in the ELAIS catalogue (Rowan-Robinson et
al. \cite{Row04}).  In the other ambiguous case a MIR
source is associated with a $z=0.1882$ optical
spectroscopic detection of a ``spiral'' galaxy in the ELAIS catalogue.
However, the optical-NIR magnitudes, colours and size of this galaxy
are that of Im type, and the strong 6.7 and 15 $\mu$m emission
detected would be difficult to understand (unless the source is a
strong starburst or AGN, which should have been obvious
from the spectrum).  Instead, there is a
very red ERO 6\arcsec \ away to which the IR SED would fit much
better (the $1\sigma$ positional uncertainty of the ELAIS MIR
catalogue is $\approx2$\arcsec). In fact, the 
preliminary ELAIS source in the source list used at the time of our NIR
observations, was closer to this ERO than the $z=0.188$ dwarf.

If we assumed that the IRTF frames are a non-biased sample of the NIR
sky, we would derive a 
lower limit of $3-8$\% for the dusty EROs in the total ERO population
(ignoring detection limits for the moment).  In fact, the
assumption is not totally incorrect, 
since only 1/3 of our frames contain a ISOCAM source (the reason being
the prelimary source list).

However, it is important to stress that the ELAIS mid-IR detection
limit of $\approx 0.7$ mJy at 15$\mu$m results in severe 
limitations.  This detection limit corresponds to, for example, a
starburst galaxy of $K=17.4$ brightness, with a SED shape of M82
located at $z=1$.  This would already be very close to the ULIRG limit
of $L_{IR} = 10^{12} \ L_{\sun}$ (see also e.g.\ Elbaz et al.\
\cite{Elb02}), and all our EROs are fainter than this in the NIR.    
As discussed in Paper~I, deeper mid-IR imaging with e.g.\ the
Spitzer/SIRTF mission, is clearly needed to detect counterparts of
those dusty EROs which are not (ultra)luminous IR galaxies.  

As for the more extreme dusty sources, an ARP220 SED-type object at
$z\approx1$ and $K<18$ would have $>1.3$ mJy at 15$\mu$m, and thus
should 
be readily detected by the ELAIS survey.  For comparison with other
dusty ERO searches (eg.\ Smail et al.\ \cite{Sm02}) this example case
extreme dusty ERO corresponds to a star formation rate of $SFR \approx
600 {\rm  \, M_{\sun} \ year^{-1}}$ (as calculated from Mann et al.\
\cite{Ma02} using total IR luminosity).
At $K<18$ we have 16 IRTF EROs, and a maximum of two of them are
detected in the mid-IR (the two ambigous cases).  Since we were
specifically targeting ISOCAM sources and thus would be biased to
dusty EROs, we can assign a strong upper limit of $<13$\% for $K<18$
EROs being strong starbursts of ULIRG type.  

Separation of more quiescent EROs using broad band colours 
is difficult because of the lack of optical detections (or $J,H$
band data) for many of the EROs.  However, we do note that for example
EROs \#15--27 in Table~\ref{Tab0-phot} are the same population as that
identified in Paper~I as early type galaxies based on their
red $r-K$ but bluer $i-K$ colours (or equivalently, red $r-i$ and blue
$i-Z$ colour, see e.g.\ SEDs in Fig.~7. of Paper~I).  Since part of
the other 
EROs are expected to also be ellipticals, the indication is totally
consistent with other studies finding the early type vs.\ dusty EROs
to be somewhere around 50\% with a large uncertainty.  We
also calculated photometric redshifts for those of our EROs which have 
photometric detections in three or more bands
using the HYPERZ code (Bolzonella, Miralles \& Pell\'o \cite{Bo00}):
These 14 EROs have redshifts ranging from 0.7 to 1.5, with an average
of $z=1.1$.  Nearly all of them, 12/14, are best fitted by an evolved
stellar population 
of at least $400$ Myr old after an instantaneous starburst, i.e.\
essentially an elliptical.

\subsection{ERO counts and surface densities}

We detect a total ERO surface density at $K<18.5$ 
of $0.26 - 0.36 \ \rm arcmin^{-2}$  
in the IRTF sample, depending on the colour selection
criteria. See Table~\ref{tab2}.  At $K \le 18$ we have $0.19 \ \rm
arcmin^{-2}$ $r-K$ 
selected EROs.  These IRTF counts are a factor of $\sim2-3$ higher than
average literature counts using the equivalent $R-K>5$, $K<18$ limits:
Daddi et al.\ (\cite{Da00})  give $0.08 \ \rm 
arcmin^{-2}$, Roche et al. 
(\cite{Ro02}) find $0.05 \ \rm arcmin^{-2}$ $R-K>5.0$  objects.  Yan
\& Thompson (\cite{Ya03}) give a compilation of 
numerous counts with different selection criteria, including $I-K>4$,
$I-H>3$, and $R-K>5$, which have an approximate average of $0.1 \
\rm arcmin^{2}$ at $K<18$. 

Some of the differences between the surface
densities of the samples may of course be attributed to different
selection criteria and different photometric measurement methods. 
Previous surveys have however found large field-to-field variations in
ERO densities.  As will be discussed in the next section, the 
apparent overdensity of our IRTF EROs results from strong
clustering of bright EROs around several ISOCAM targets.

   \begin{table*}
\begin{center}
\begin{tabular}{lrccrccrcc}
     \hline
     \hline
     \noalign{\smallskip}
             &   \multicolumn{3}{c}{$r-K>5.5$} & \multicolumn{3}{c}{$r-K>5.8$} & \multicolumn{3}{c}{$i-K>4.4$} \\
            \noalign{\smallskip}
 \multicolumn{10}{c}{All frames (area 74 arcmin$^{2}$)} \\
$K$ limit &  & $\Sigma_{K}$ &   &  & $\Sigma_{K}$ &   &
 &    $\Sigma_{K}$ &    \\
(mag)     &  N & $\rm arcmin^{-2}$ &   & N & $\rm arcmin^{-2}$ &   &
     N & $\rm arcmin^{-2}$ &   \\
            \noalign{\smallskip}
            \hline
            \noalign{\smallskip}
$K \leq 17.5$  & 3  & 0.041 &  &  2 & 0.027 &  &  2 & 0.027 &  \\
$K \leq 18.0$  & 14 & 0.189 &  & 11 & 0.149 &  &  9 & 0.122 &  \\
$K \leq 18.5$  & 27 & 0.365 &  & 12 & 0.162 &  & 19 & 0.257 &  \\
$K \leq 18.7$  & 27 & 0.365 &  & 12 & 0.162 &  & 27 & 0.365 &  \\
\noalign{\smallskip}
\hline
\noalign{\smallskip}
\multicolumn{10}{c}{ISO frames (area 17.2 arcmin$^{2}$) }   \\
 &  & $\Sigma_{K}$ &  & & $\Sigma_{K}$ &  & &
     $\Sigma_{K}$ &   \\
     &  N & $\rm arcmin^{-2}$ & $\sigma_{\rm exc}$ & N & $\rm arcmin^{-2}$ & $\sigma_{\rm exc}$ &
     N & $\rm arcmin^{-2}$ & $\sigma_{\rm exc}$ \\
\noalign{\smallskip}
            \hline
            \noalign{\smallskip}
$K \leq 17.5$  & 2  & 0.116 & ---  & 1 & 0.058 & --- & 1 & 0.058 & ---
            \\
$K \leq 18.0$  & 8 & 0.465 & 3.4 & 6 & 0.349 & 3.1 & 3 & 0.174 & 0.6 \\
$K \leq 18.5$  & 13 & 0.756 & 2.6 & 7 & 0.407 & 3.8 & 7 & 0.407 & 1.1
            \\
$K \leq 18.7$  & 13 & 0.756 & 2.9 & 7 & 0.407 & 3.7 & 12 & 0.698 & 2.3 \\
\noalign{\smallskip}
\hline
\noalign{\smallskip}
\multicolumn{10}{c}{field frames (area 56.4 arcmin$^{2}$)} \\
   &  & $\Sigma_{K}$ &  & & $\Sigma_{K}$ &  & &
     $\Sigma_{K}$ &   \\
     &  N & $\rm arcmin^{-2}$ &  & N & $\rm arcmin^{-2}$ &  &
     N & $\rm arcmin^{-2}$ &  \\
            \noalign{\smallskip}
\hline
            \noalign{\smallskip}
$K \leq 17.5$  & 1  &  0.018 & & 1 &  0.018 & & 1   &  0.018 \\
$K \leq 18.0$  & 6 &   0.106 & & 5 &  0.089 & & 6   &  0.106 \\
$K \leq 18.5$  & 14 &   0.248 & & 5 &  0.089 & & 12   & 0.213 \\
$K \leq 18.7$  & 14 &   0.248 & & 5 &  0.089 & & 15  & 0.266 \\
\noalign{\smallskip}
\hline
\end{tabular}
\end{center}
\caption[]{The sample of IRTF EROs.  Cumulative counts are given
for different ERO colour selection criteria. Additionally, counts are
given separately for the total survey, those areas with a faint
mid-IR source within 90\arcsec, and those without.  The latter counts
are intended to be taken as field ERO counts. For the ISO related
counts we also tabulate the significance of the excess of EROs in the
corresponding magnitude and colour bins.}
\label{tab2}
\end{table*}

\subsection{ERO excess around ISOCAM fields}
\label{excess}

The ERO distribution is not uniform, qualitatively there are clearly
more EROs in IRTF frames which also have an ISOCAM galaxy.  The
contrast is even larger when only faint MIR sources are considered.
 To obtain the result quantitavely, we calculated the surface
densities of EROs within 90\arcsec \ radii around those $15\mu$m
ISOCAM 
sources which had spectroscopic or photometric redshifts above
$z>0.5$ or which did not have redshift estimates (i.e.\ the
faintest NIR cases with no optical counterparts to WFS limits).  The
search radius was selected to 
roughly match sizes of 
central regions of rich clusters at $z\sim 1$, where 90\arcsec \
corresponds to $\sim 0.8$ Mpc.  Cases where an ISOCAM source fell
outside the IRTF frame but 
closer than 90\arcsec \ from our NIR data, were included
for the appropriate part of the frame area.   
The nearby, $z<0.5$ ISO-galaxies were excluded and added to the
field sample, with one exception: we chose to assign source
ELAISC15\_J163541.7+405900, the se\-cond ``ambiguous'' source in
Sect.~\ref{excess}, for reasons discussed therein, to an ERO 6\arcsec
\ away with a
photometric redshift of $z=0.85$ instead of the official ELAIS
catalogue identification of a
$z=0.188$ spiral.  The main results of this paper do not change with
this decision, and results will also be given for the case with a
contrary decision.  The relevant ISO sources are listed in
Table~\ref{tab4} with the corresponding number of detected EROs.

For the 20 ISOCAM source related EROs in an area of 17.2 
arcmin$^2$, we get a surface density of $1.16 \ \rm arcmin^{-2}$,
whereas for the ``field ERO'' sample, N=20 in 54.6 $\rm arcmin^{2}$, 
the surface density is $0.37 \ \rm arcmin^{-2}$.  We calculated
the densities in true non-MIR frames and nearby-MIR frames also
separately: the results are 0.35 (N=14) and 0.38 (N=6) EROs per
arcmin$^{2}$, respectively.  Changing the category of the 4 EROs found
around ELAISC15\_J163541.7+405900, the ISO-source ERO density would
decrease 
to 1.06 arcmin$^{2}$ and the field density increase to $0.43 \ \rm
arcmin^{-2}$.

\subsubsection{The EROs in ISO fields}

Figure~\ref{erocounts2} shows the cumulative IRTF
counts separated into the ISOCAM source related EROs (triangles) and
field EROs (circles).  

We note that the surface density
measured from the ISOCAM frames is fully consistent with ERO
surveys performed in fields of high-$z$ AGN, radio galaxies, or
radio-loud QSO (eg.\ Wold et al.\ \cite{Wo03}; Cimatti et
al. \cite{Ci00}; Hall \& Green \cite{Ha98}).
With this in mind, it is interesting to test how our EROs
correlate with the ELAIS 
final catalogue radio-sources.  We searched for EROs around
(45\arcsec \ radius used) VLA 1.4 GHz detections with spectroscopic
or photometric redshifts of 
$z>0.5$.  The ERO density becomes
$0.80\ \rm arcmin^{-2}$, twice as high as the ``field'' value. 
We note that 2 of the 8 ISOCAM sources listed in Table~\ref{tab4} are
also radio sources; also, 3 other listed cases have a radio source in
the searched area. Thus, there clearly is no one-to-one corresponce
of radio and mid-IR detections, though there is overlap.  Further
study of the relation of the mid-IR sources to the VLA sources is out
of the scope of the present paper.

\subsubsection{The field EROs}

The derived field ERO 
surface density is totally consistent with other field ERO surveys.
The surface density of EROs provides a constraint for passive
evolution and hierarchical galaxy formation models.  However, since
the present 
survey does not add significantly to the covered area of large random
field ERO surveys at the depth of our survey, we do not
discuss the implications of surface density in detail here.  
Merely for reference, we plot a model from
Daddi et al.\ (\cite{Da00b}) in Fig.~\ref{erocounts2}, which predicts
the number of EROs resulting from a single $e$-folding time $\tau =
0.1$  Gyr starburst at $z=3$ and pure luminosity evolution since.  It
is seen that our IRTF counts, along with brighter counts from our
Paper~I and literature counts in the range $K=17-20$ fit this PLE model
remarkably well.

\begin{table}
\begin{center}
\footnotesize
\begin{tabular}{llcl}
     \hline
     \hline
     \noalign{\smallskip}
 & ISO Object  & N EROs &  Redshift \\  
            \noalign{\smallskip}
            \hline
            \noalign{\smallskip}
1 & ELAISC15\_J163417.9+405653 & 2 & 1.0 \\ 
2 & ELAISC15\_J163515.6+405608 & 0 & 2.5 \\ 
3 & ELAISC15\_J163531.1+410025 & 4 & 1.150 \\  
4$^a$ & ELAISC15\_J163541.7+405900 & 4 & 0.9 \\ 
5 & ELAISC15\_J163543.1+410750 & 4 & -- \\ 
6 & ELAISC15\_J163615.7+404759 & 3 & 0.9 \\ 
7 & ELAISC15\_J163655.8+405909 & 2 & 2.610 \\ 
8 & ELAISC15\_J163730.9+410447 & 1 & -- \\ 
            \noalign{\smallskip}
            \hline
            \noalign{\smallskip}
9$^b$ & ELAISC15\_J160718.9+544406 & 0 & -- \\ 
10 & ELAISC15\_J160741.1+550152 & 1 & 0.3 \\ 
11 & ELAISC15\_J161013.4+550648 & 1 & 0.4 \\ 
12 & ELAISC15\_J161101.3+543331 & 0 & 0.235 \\ 
13 & ELAISC15\_J161103.0+543220 & 0 & 0.147 \\ 
14 & ELAISC15\_J163359.3+410917 & 0 & 0.137 \\ 
15 & ELAISC15\_J163414.2+410317 & 4 & 0.040 \\
16 & ELAISC15\_J163629.1+411441 & 0 & 0.4 \\
17 & ELAISC15\_J163640.0+405538 & 0 & 0.106 \\ 
\noalign{\smallskip}
\hline
\end{tabular}
\end{center}
\caption[]{List of ELAIS 15 $\mu$m sources associated with our IRTF
  frames.  The number of EROs within 90\arcsec \ from the mid-IR
  source is 
  given.  Also, we list the spectroscopic (3 decimals) or best-fitting
  photometric (1 decimal) redshift for each source
  from Rowan-Robinson et al.\ (\cite{Row04}). The sources from \#9
  onward are those which have redshift $z<0.5$ and were excluded from
  the proper ``ISO-related'' frames from which the overdensities are
  derived.  The IRTF frames corresponding to all
  the listed sources contain an area 
  of 33.1 arcmin$^2$.  $^a$Note: ELAIS catalogue associates this
  object with a $z=0.189$ spiral, we associated it with a
  higher redshift alternative, see text. $^b$Note: optical data
  fall in CCD gap, thus no redshift estimate; clearly a nearby galaxy
  pair from NIR data.} 
\label{tab4}
\end{table}

   \begin{figure}
\resizebox{9.0cm}{!}{\includegraphics{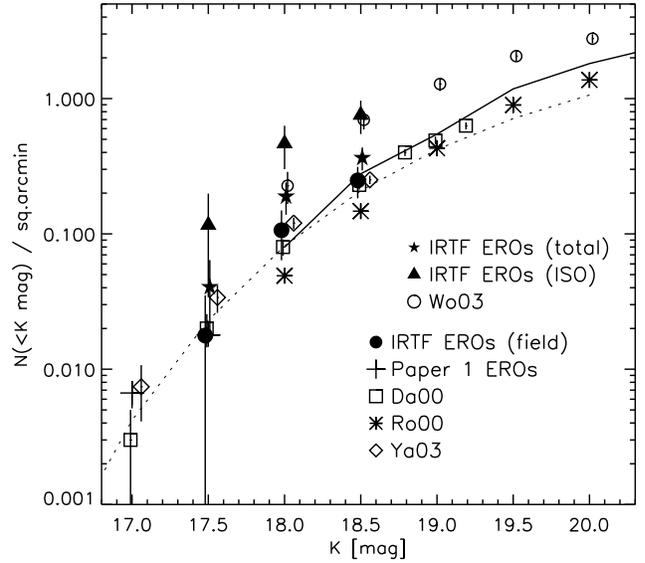}}
      \caption[]{Cumulative ERO counts of the IRTF
	survey for the $r-K>5.5$ selection (solid symbols): the stars
	show the total IRTF ERO counts, the triangles counts around
	faint ISOCAM sources, and the circles the ``field'' frame
	counts, as discussed in text.  Error bars are Poissonian.  
 The recent ERO counts in the fields of radio-loud QSOs by Wold et al.\
	(\cite{Wo03}) are shown as open circles and are seen to agree
	very well with our faint mid-IR related ERO counts. 
	Literature field ERO counts with $R-K>5$ from Daddi et al.\
	(\cite{Da00}) and Roche et al.\ (\cite{Ro02}) are shown, as
	well as the $i-K>4$ EROs of Yan \& Thompson (\cite{Ya03}). The
	thick solid line is an averaged compilation of other field ERO
	counts by Wold et al.\ (\cite{Wo03}). The dotted line shows
	for reference a 
	best fitting PLE model (Daddi et al.\ \cite{Da00b}) discussed
	in Paper~I.} 
         \label{erocounts2}
   \end{figure}

\subsection{Significance of the ERO overdensity}

Our IRTF sample is non-continuous in sky coverage and we are unable to
perform quantitative clustering analysis with the sample.
However, since EROs are found to be strongly clustered in random field
surveys (e.g.\ Daddi et al. \cite{Da00}; Roche et al. \cite{Ro02}), it
is important to estimate the significance of the excess surface
densities found above.  We follow here the 
treatment of Wold et al.\ (\cite{Wo03}).  The significance of the
excess can be given by

\begin{equation}
\sigma_{exc} = (N_{\rm MIR} - N_{\rm field}) / \sigma
\label{eq1}
\end{equation}
where $N_{\rm MIR}$ is the total number of EROs in the fields of
ISOCAM mid-IR galaxies, and $N_{\rm field}$ is the number in the field
IRTF frames.  The expected field count fluctuations are given by
$\sigma$ and can be calculated from 

\begin{equation}
\sigma^{2} = N_{\rm field}( 1 + N_{\rm field} A_{\omega} C )
\label{eq2}
\end{equation}
where $A_{\omega}$ is the amplitude of the field ERO angular two-point
correlation function and $A_{\omega}C$ the ``integral constraint''
necessary to correct for the finite survey area (see e.g.\ Daddi et
al.\ \cite{Da00}; Roche et al. \cite{Ro02}).  
We use the correlation amplitude $A_{\omega}$ values from
the large area ERO clustering 
survey of Daddi et al.\ (\cite{Da00}) which matches well our
magnitude range of $K<18.5$, and is consistent at the faint end with 
other 
determinations (Roche et al.\ \cite{Ro02, Ro03}; Firth et al.\
\cite{Fi02}).  In addition, Daddi et al.\ (\cite{Da00}) find that the
following 
approximation, which we also adopt, fits the actual detected
clustering: $C \approx 58 \times {\rm Area}^{-0.4}$, where the area is
expressed in arcmin$^2$.

Our survey area for
the mid-IR frames is 17.2 arcmin$^2$ and a total number of 20 EROs are
found in the area, compared to 20 field EROs in 56.4 arcmin$^2$: using
the equations above we calculate a significance for the excess
$\sigma_{\rm exc} = 3.0$.  

To obtain a measure of quantitative uncertainty in the
  overall $\sigma_{\rm exc}$ value just calculated, we performed a
  simple Monte Carlo
  bootstrap analysis (see e.g.\ Press et al.\ \cite{Pr92}): Each of
  the N=20 MIR-related EROs and the N=20 field 
  EROs is replaced by a randomly selected ERO from the original
  sample; the new samples are typically smaller, since the any given
  object may be picked upto N=20 times from the original sample.  The
  process is repeated 
  numerous times, and the calculation using Equations~1 and~2 is
  repeated after each round.  We find a standard deviation of 0.59 for
  $10^5$ 
  simulated $\sigma_{\rm exc}$ values (the average becomes 2.82, i.e.\
  essentially the same as the initial value of 3.0).  We thus conclude
  that the overdensity of EROs -- in the ultimately quite small area
  of 17.2 arcmin$^2$ of the ISOCAM detection related IRTF frames -- is
  statistically significant. 

Significances  $\sigma_{\rm exc}$ for other magnitude cut-offs and
separated for $r-K$ and $i-K$ selections, are given in
Table~\ref{tab2}.  The raw numbers of objects are becoming small, and
the confidences thus lower.  However, we do note that a stronger
clustering around the ISOCAM sources is suggested for  $r-K$ EROs than
for $i-K$ selected EROs.  The $i-K$ EROs
have higher surface densities in the ISOCAM field by factors of
$1.5-2.5$ compared to field frames, whereas the $r-K$ selected
overdensities are by factors of $3-5$.  If correct, and not only
result of small number statistics, the effect would be consistent with
the picture where the $r-K$ EROs are preferentially early
types, and thus more clustered (Daddi et al. \cite{Da02}), and the
$i-K$ EROs are more biased to dusty starbursts and/or those galaxies
with continuing star formation.

\subsection{Radial distribution}

We attempt to determine the
distribution of the EROs making up the overdensity as a function of
distance from the corresponding ISOCAM source.  Fig.~\ref{radial}
shows the surface densities of EROs calculated using the appopriate
available areas around the ISOCAM sources.  The range $0-30$\arcsec,
which corresponds to radial distance of 250 kpc at $z=1$ includes
12 out of the 20 EROs.  It is seen that the first
bin, corresponding to 6 EROs closer than 15\arcsec \ from an ISOCAM
galaxy, produces a very high surface density.  Mean surface
densities of the ISO related and field frames are also indicated.
The trend is similar to that found for EROs by Wold et al.\
(\cite{Wo03}) at slightly fainter magnitudes around high-$z$ quasars,
and by Best et al.\ 
(\cite{Be03}) for red $R-K>4$ galaxies around radio-loud AGN.
We simulated our radial distribution again with the bootstrap
  method, 
by replacing the original 20 EROs, with given radial distances, by
randomly drawn EROs from the same sample: the error bars in
Fig.~\ref{radial} reflect the standard deviation of $10^5$
si\-mu\-lated  
surface densities in each bin. 
However, we note that excluding object \#4 in Table~\ref{tab4} would
make the radial distribution flatter and significant only at
$\sim2\sigma$ level in the two first bins. 

   \begin{figure}
\resizebox{8.5cm}{!}{\includegraphics{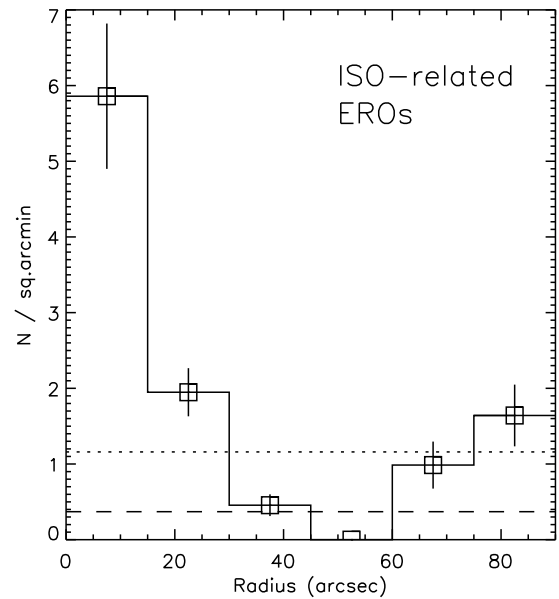}}
      \caption[]{Average radial distribution of EROs in the areas
	related to an ISOCAM source.  The mean {\em field} ERO surface
	density in our survey is plotted with the dashed line, and the
	density of ISO-frame EROs averaged over the whole distance
	range to 90\arcsec \ is shown with the dotted line. The
	error bars 
	are derived from a bootstrap analysis (see text) -- there is 
	a significant concentration of EROs within 30\arcsec of the
	ISOCAM galaxy. 
}
         \label{radial}
   \end{figure}

\subsection{High-$z$ clusters of galaxies?}
\label{clusters}

We have demonstrated that fields containing faint mid-IR galaxies
have a significant overdensity of EROs compared to field populations of
EROs.  Without spectra it is impossible to definitely show that 
the EROs are related to the mid-IR galaxy in question.  Since most
EROs have magnitude upper limits only, it is not possible to do
colour-magnitude sequence analysis either.  The concentrated radial
distribution nevertheless suggests a physical correlation with the
EROs and the central mid-IR source.  An alternative for a
physical connection would be that the EROs belong to a foreground
structure which enhances the probability of detection of background
objects through gravitational lensing -- see discussion eg.\ in
Cimatti et al.\ (\cite{Ci00}) and Wold et al.\ (\cite{Wo03}).

If the EROs indeed trace a cluster of galaxies, whether at
the mid-IR source redshift or in the foreground, one would expect a
larger fraction of EROs in a $K$-selected sample of galaxies at that
region, since clusters are known to have a larger fraction of
ellipticals than the field (see e.g.\ Dressler \cite{Dr80}). From
Table~\ref{tab5} it is evident that this indeed is the case for our
ISO related IRTF fields.  ERO fractions of $\sim20$\% are very similar
to findings of Wold et al.\ (\cite{Wo03}) in their $z\sim2$ QSO
fields. The table also lists the ERO fractions in our whole survey.
These values fall consistently in between the high values of the
mid-IR fields and
those in random fields as in Daddi et al.\ (\cite{Da00}) and our
Paper~I values, which have fractions increasing from 0.01 to 0.08 in
the tabulated magnitude range.

\begin{table}
\begin{center}
\begin{tabular}{lccc}
     \hline
     \hline
     \noalign{\smallskip}
             &   $r-K>5.5$ & $r-K>5.8$ & $i-K>4.4$ \\
     \hline
     \noalign{\smallskip}
\multicolumn{4}{c}{ISO frames} \\
     \noalign{\smallskip}
$K < 17.5$  &  0.09 & 0.09 & 0.09 \\
$K < 18.0$  &  0.26 & 0.19 & 0.10 \\
$K < 18.5$  &  0.21 & 0.14 & 0.11 \\
$K < 18.7$  &  0.20 & 0.14 & 0.18 \\
     \noalign{\smallskip}
     \hline
     \noalign{\smallskip}
\multicolumn{4}{c}{All frames} \\
     \noalign{\smallskip}
$K < 17.5$  &  0.03 &  0.02 & 0.02 \\ 
$K < 18.0$  &  0.09 &  0.07 & 0.06 \\ 
$K < 18.5$  &  0.12 &  0.07 & 0.08 \\ 
$K < 18.7$  &  0.11 &  0.05 & 0.10 \\ 
     \noalign{\smallskip}
     \hline
\end{tabular}
\end{center}
\caption[]{ERO fractions among $K$-selected galaxies.}
\label{tab5}
\end{table}

\section{Summary}

We have searched for EROs in a targeted near-IR survey 
in ELAIS fields, using $r-K\ge5.5$, $r-K\ge5.8$, and $i-K\ge4.4$
colour criteria. 
These are equivalent to the commonly used $R-K\ge5$, $R-K\ge5.3$, and
$I-K\ge4$.  We find 40 EROs, 14 of which are common to both
$r-K\ge5.5$  
and $i-K\ge4.4$ selection.  We study the number counts and properties
of these EROs, both close to and away from the mid-IR ELAIS sources,
and summarize the results as follows:

\begin{itemize}

\item Taking advantage of overlapping mid-IR data, we search for dusty
  EROs, 
since only these should be detected with the used $15\mu$m ISOCAM
band.  One clear mid-IR ERO is found, and two other which are
somewhat ambiguous. Taking into account detection limits we can limit
  the number of strong ULIRG type starbursts in the $K<18$ ERO
  population to $<13$\%.

\item Broad-band colours and SED fits of those EROs with
  detections in 
  at least 3 bands suggest these are mainly early type evolved
  galaxies at $z\sim1$.

\item Cumulative number counts are provided for the EROs.  
The total counts are
found to be higher than field counts in the literature, but this is
due to excess EROs discovered in ISOCAM fields.  When fields with no
mid-IR sources are considered, the ERO counts fit well other data.
Our field ERO counts, as well as other literature data, are very well
fit by pure luminosity evolution models in our magnitude range.

\item We derive a significant overdensity of EROs in areas within
  90\arcsec \ of faint $z>0.5$ mid-IR sources.  The result is
  equivalent to what other studies have found in high redshift AGN and
  QSO fields. 
  The high fraction of EROs among all $K$-selected galaxies in these
  fields suggests that the EROs are part of high redshift
  clusters of galaxies. The fields are small, however, and clusters and
  e.g.\ filaments of large scale structure cannot be properly
  differentiated.  Furthermore, $r-K$ selected EROs at $K<18.5$
  are (marginally) more clustered in the mid-IR fields than the $i-K$
  selected EROs, consistent with a picture that the
  $r-K$ selected EROs contain preferentially more early type galaxies
  than the $i-K$ EROs.

\item It is not yet clear whether the ERO overdensities are physically 
connected to the mid-IR sources (or AGN) or rather part of a foreground
  cluster or other surface density peak grationally enhancing the
  detections of background mid-IR sources.  However, the radial
  distribution of EROs around the mid-IR detection suggests a physical
  connection.

\end{itemize}

\begin{acknowledgements}

We thank Eric V.\ Tollestrup for valuable help with the IRTF
observations.  We acknowledge the work done by the ELAIS collaboration
and ISO science centre in bringing together the
ELAIS products. We wish to thank Eduardo
Gonzalez-Solares for help with the WFS photometric catalogue and
Margrethe Wold, Emanuele Daddi, Kalevi Mattila, and an anonymous
referee for useful comments. 

\end{acknowledgements}


\begin{thebibliography}{}

\bibitem[2003]{Al03} Almaini, O., Scott, S.E., Dunlop, J.S. et al.
	2003, MNRAS, 338, 303

\bibitem[2002]{Ba02} Basilakos, S., Georgantopoulos, I.,
	P\'erez-Fournon, I. et al., 2002, MNRAS, 331, 417

\bibitem[2003]{Ba03} Bahcall, N., McKay, T.A., Annis, J., et al.\
  2003, ApJS, 148, 243

\bibitem[1996]{Be96} Bertin, E., Arnouts, S., 1996,
	A\&AS, 117, 393


\bibitem[2003]{Be03} Best, P.N., Lehnert, M.D., Miley, G.K.,
  R\"ottgering, H.J.A. 2003, MNRAS, 343, 1


\bibitem[2000]{Bo00} Bolzonella, M., Miralles, J.-M., Pell\'o, R.,
	2000 A\&A, 363, 476


\bibitem[1996]{Ce96} Cesarsky, C.J., Abergel, A., Agnese, P., et al.
  1996, A\&A, 315, L32


\bibitem[2000]{Cha00} Chapman, S.C., McCarthy, P.J., Persson, S.E
  2000, AJ, 120, 1612


\bibitem[2000]{Ci00} Cimatti, A., Villani, D., Pozzetti, L, di Serego  
Alighieri, S. 2000, MNRAS, 318, 453

\bibitem[2002]{Ci02} Cimatti, A., Daddi, E., Mignoli, M. et al. 2002,
	A\&A, 381, L68

\bibitem[2003]{Ci03} Cimatti, A., Daddi, E., Cassata, P. et al. 2003,
  A\&A, 412, L1


\bibitem[1994]{Co94} Cohen, M. 1994, AJ, 107, 582 

\bibitem[2000]{Co00} Cole, A., Lacey, C.G., Baugh, C.M., Frenk, C.S.
  2000, MNRAS, 319, 204


\bibitem[2000a]{Da00} Daddi, E., Cimatti, A., Pozzetti, L. et al.
	2000, A\&A, 361, 535

\bibitem[2000b]{Da00b} Daddi, E., Cimatti, A., Renzini, L. 2000,
	A\&A, 362, L45

\bibitem[2002]{Da02} Daddi, E,. Cimatti, A., Broadhurst, T., et al.
  2002, A\&A, 384, L1

\bibitem[2002]{Dah02} Dahn, C.C., Harris, H.C., Vrba, F.J., et al.
  2002, AJ, 124, 1170

\bibitem[1995]{De95} Dey, A., Spinrad, H., Dickinson, M. 1995,
	ApJ, 440, 515


\bibitem[1980]{Dr80} Dressler, A. 1980, ApJ, 236, 351


\bibitem[2000]{Ef00} Efstathiou, A., Oliver, S., Rowan-Robinson, M. et
  al., 2000, MNRAS, 319, 1169

\bibitem[1962]{Eg62} Eggen, O.J., Lynden-Bell, D. \& Sandage, A.R.
	1962, ApJ, 136, 748

\bibitem[2002]{Elb02} Elbaz, D., Cesarsky, C.J., Chanial, P. et al. 2002,
	A\&A, 384, 848


\bibitem[2002]{Fa02} Fadda, D., Flores, H., Hasinger, G., et al. 2002,
  A\&A, 383, 838 

\bibitem[2002]{Fi02} Firth, A.E., Somerville, R.S., McMahon, R.G. et
	al. 2002 MNRAS, 332, 617


\bibitem[1996]{Fu96} Fukugita, M., Ichikawa, T., Gunn, J.E. et
	al. 1996, AJ, 111, 1748

\bibitem[2003]{Gi03} Gilbank, D.G., Smail, I., Ivison, R.J., Packham,
  C. 2003, MNRAS, 346, 1125

\bibitem[2001]{Go01} Gonzalez, A.H., Zaritsky, D., Dalcanton, J.J.,
  Nelson, A. 2001, ApJS, 137, 117

\bibitem[2004]{Go04} Gonzalez-Solares, E.A., Perez-Fournon, I.,
  Rowan-Robinson, M. et al. 2004, MNRAS, submitted (astro-ph/0402406)


\bibitem[1998]{Ha98} Hall, P.B., Green, R.F. ApJ, 507, 558

\bibitem[2001]{Ha01} Hall, P.B., Sawicki, M., Martini, P., et al.\
  2001, AJ, 121, 1840


\bibitem[1994]{Hu94} Hu, E.M., Ridgway, S.E. 1994,
	AJ, 107, 1303

\bibitem[2001]{Ir01} Irwin, M., Lewis, J. 2001,
	NewAR, 45, 105

\bibitem[1992]{Ja92} Jarrett, T.H. 1992, PhD thesis,  University of
  Massachusetts, Amherst 


\bibitem[1980]{Kr80} Kron, R.G. 1980,
	ApJS, 43, 305

\bibitem[2004]{La04} La Franca, F., Gruppioni, C., Matute, I., et
  al. 2004, AJ, 127 (in press)

\bibitem[1992]{La92} Landolt, A.U. 1992, AJ, 104, 340   

\bibitem[1975]{La75} Larson, R.B. 1975,
	MNRAS, 173, 671


\bibitem[2002]{Ma02} Mann, R.G, Oliver, S., Carballo, R., et al. 2002,
  MNRAS, 332, 549

\bibitem[2003]{Ma03} Manners, J.C., Johnson, O., Almaini, O., et al.
  2003, MNRAS, 343, 293 

\bibitem[2001]{Mc01} McMahon, R.G., Walton, N.A., Irwin, M.J. et al.
	2001, NewAR, 45, 97

\bibitem[1992]{Mc92} McCarthy, P.J., Persson, S.E., West, S.C.
	1992, ApJ, 386, 52


\bibitem[2002]{Mo02} Mohan, N.R, Cimatti, A., R\"ottgering, H.J.
  2002, A\&A, 383, 440

\bibitem[2004]{Mo04} Moustakas, L.A., Casertano, S., Conselice, C. et
  al. 2004, ApJ, 600, L131

\bibitem[2003]{Mu03} Mullis, C.R., McNamara, B.R., Quintana, H., et
al. 2003, ApJ, 594, 154

\bibitem[2000]{Ol00} Oliver, S., Rowan-Robinson, M. et al. 2000,
	MNRAS, 316, 749

\bibitem[1999]{Ol99} Olsen, L.F., Scodeggio, M., Da Costa, L. et al.\
  1999, A\&A, 345, 363


\bibitem[1992]{Pr92} Press, W.H., Teukolsky, S.A., Vetterling, W.T.,
  Flannery, B.P. 1992, Numerical recipes in C, The Art of Scientific
  Computing, Second Edition, Cambridge University Press, p.\ 691

\bibitem[2002]{Ro02} Roche, N.D., Almaini, O., Dunlop, J.S., Ivison,
	R.J., Willott, C.J. 2002, MNRAS, 337, 1282

\bibitem[2003]{Ro03} Roche, N.D., Dunlop, J.S., Almaini, O. 2003,
  MNRAS, 346, 803 


\bibitem[2004]{Row04} Rowan-Robinson, M., Lari, C., Perez-Fournon, I.,
  et al., 2004, MNRAS, in press (astro-ph/0308283)

\bibitem[2002]{Sa02} S\'anchez, S.F., Gonz\'alez-Serrano, J. I. 2002,
  A\&A, 396, 773 


\bibitem[2002]{Sc02} Scott, S.E., Fox, M.J., Dunlop, J.S. et al.
  2002, MNRAS, 331, 817

\bibitem[2000]{Se00} Serjeant, S., Oliver, S., Rowan-Robinson, M. et
  al. 2000, MNRAS, 316, 768

\bibitem[1998]{Si98} Silva, L., Granato, G.L., Bressan, A., Danese,
        L. 1998, ApJ, 509, 103


\bibitem[2002]{Sm02} Smail, I., Owen, F.N., Morrison, G.E., et al.
	2002, ApJ, 581, 844


\bibitem[1999]{So99} Somerville, R.S., Primack, J.R. 1999, MNRAS,
  310, 1087

\bibitem[2004]{So04} Somerville, R.S., Moustakas, L.A., Mobasher, B.
  2004, ApJ, 600, L135

\bibitem[2003]{Ta03} Takata, T., Kashikawa, N., Nakanishi, K., et al.
  2003, PASJ, 55, 789

\bibitem[2004]{Vac04} Vaccari, M., Lari, C., Angeretti, L., et
  al. 2004, MNRAS, submitted (astro-ph/0404315)

\bibitem[2000]{Va00} V\"ais\"anen, P., Tollestrup, E.V., Willner,
	S.P., Cohen, M. 2000, ApJ, 540, 593

\bibitem[2004]{Va04} V\"ais\"anen, P., Johansson, P.H. 2004, A\&A,
	in press (astro-ph/0404129) (Paper~I)

\bibitem[2004]{We04} Webb, T.M.A, Brodwin, M., Eales, S., Lilly,
  S.J. 2004, ApJ, in press (astro-ph/0311598)

\bibitem[2002]{We02} Wehner, Barger, A., Kneib, J.-P. 2002, ApJ, 577,
  83 

\bibitem[1978]{Wh78} White, S.D.M., Rees, M.J. 1978, 
        MNRAS , 183, 341

\bibitem[1991]{Wh91} White, S.D.M., Frenk, C.S. 1991,
	ApJ, 379, 52

\bibitem[2003]{Wo03} Wold, M., Armus, L., Neugebauer, G., Jarrett,
  T.H., Lehnert, M.D. 2003, AJ, 126, 1776

\bibitem[2003]{Ya03} Yan, L., Thompson, D. 2003, ApJ, 586, 765



\end{thebibliography}
\end{document}